\pdfoutput=1

\PassOptionsToPackage{dvipsnames}{xcolor}

\documentclass[11pt]{article}

\usepackage[]{acl}

\usepackage{times}
\usepackage{latexsym}

\usepackage[T1]{fontenc}

\usepackage[utf8]{inputenc}

\usepackage{microtype}

\usepackage{booktabs}
\usepackage{subcaption}
\usepackage{makecell}
\usepackage{graphicx}
\usepackage{amsmath}
\usepackage{pifont}
\usepackage{tikzsymbols}
\usepackage{xcolor}
\newcommand{\xmark}{{\color{red} \ding{55}}}%
\newcommand{\cmark}{{\color{green} \ding{51}}}

\title{Improving Text Matching in E-Commerce Search with \\ A Rationalizable, Intervenable and Fast Entity-Based Relevance Model}

\author{ Jiong Cai$^\diamond$, Yong Jiang$^\dagger$, Yue Zhang$^\dagger$, Chengyue Jiang$^\diamond$, Ke Yu$^\dagger$, Jianhui Ji$^\dagger$, Rong Xiao$^\dagger$,\\ \textbf{Haihong Tang$^\dagger$, Tao Wang, Zhongqiang Huang$^\dagger$, Pengjun Xie$^\dagger$, Fei Huang$^\dagger$, Kewei Tu$^\diamond$} \thanks{\hspace{1mm} Yong Jiang and Kewei Tu are the corresponding authors. This work was conducted when Jiong Cai was interning at Alibaba DAMO Academy. }  \\
$^\diamond$School of Information Science and Technology, ShanghaiTech University \\
$^\dagger$DAMO Academy, Alibaba Group \\
 {\tt \{caijiong,jiangchy,tukw\}@shanghaitech.edu.cn} \\ 
 {\tt \{yongjiang.jy,shiyu.zy\}@alibaba-inc.com}
}

\begin{document}
\maketitle
\begin{abstract}
Discovering the intended items of user queries from a massive repository of items is one of the main goals of an e-commerce search system. Relevance prediction is essential to the search system since it helps improve performance. When online serving a relevance model, the model is required to perform fast and accurate inference. Currently, the widely used models such as Bi-encoder and Cross-encoder have their limitations in accuracy or inference speed respectively. In this work, we propose a novel model called the Entity-Based Relevance Model (EBRM). We identify the entities contained in an item and decompose the QI (query-item) relevance problem into multiple QE (query-entity) relevance problems; we then aggregate their results to form the QI prediction using a soft logic formulation. The decomposition allows us to use a Cross-encoder QE relevance module for high accuracy as well as cache QE predictions for fast online inference. Utilizing soft logic makes the prediction procedure interpretable and intervenable. We also show that pretraining the QE module with auto-generated QE data from user logs can further improve the overall performance. The proposed method is evaluated on labeled data from e-commerce websites. Empirical results show that it achieves promising improvements with computation efficiency. 
\end{abstract}

\section{Introduction}

\begin{figure*}[t]
    \centering
    \resizebox{.8\textwidth}{!}{
     \includegraphics[height=2.8cm]{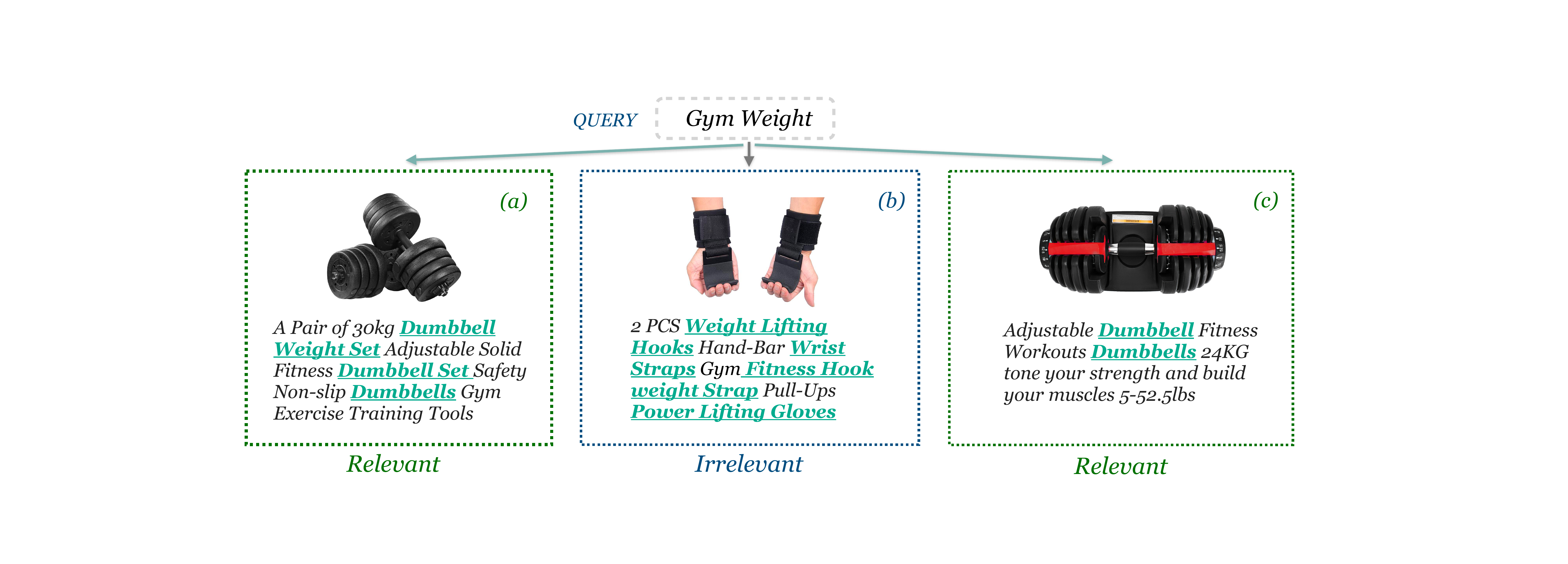}
    }
    \caption{Examples of relevance prediction. The query is \emph{"gym weight"} and items are listed with their titles. The extracted entities in the titles are underlined and colored.  The examples (a) and (c) are relevant to the query, but (b) is not.}
    \label{fig:example}
\end{figure*}

Nowadays, e-commerce platforms become a major shopping channel in people's life. The top e-commerce shopping platforms (such as Amazon, Taobao, and eBay) have hundreds of millions of items and active users. From the extremely large amounts of items, an e-commerce search system helps users find the ones they want. To achieve this goal, a relevance model is designed to measure the relevance score of the user query and item \cite{dssm-2013, DRMM-16, KNRM-2018}. Accurate relevance measurement is crucial for an e-commerce search system since displaying irrelevant items for the query would degrade the user experience and harm retention. For queries, users use short and ambiguous text to describe their search intention. On the other side, vendors tend to write long item titles with redundant phrases and do not follow any specific structures. We present a few examples in Figure \ref{fig:example}. The discrepancy between queries and item titles results in difficulties in building a fast and accurate relevance model. 

Term-based methods like BM25 \cite{robertson1995okapi} and TF-IDF are widely used to estimate the relevance in the search system. These methods utilize term frequency to measure the relevance, but they would suffer from the vocabulary difference between queries and item titles. For example, they would falsely predict relevance for case (b) and irrelevance for case (c) in Figure~\ref{fig:example}. To alleviate the problem, neural-based methods \cite{dssm-2013,shen2014learning, duet-2017, relevance-emb-2017, KNRM-2017, KNRM-2018, INR-061} map the queries and items to vectors in a dense semantic space and measure the relevance accordingly. Recently, using deep contextualized representations from Transformer-based models, such as BERT \cite{devlin2019bert}, has been proven effective in information retrieval tasks \cite{dai2019deeper, qiao2019understanding, cedr-2019, nogueira2019passage}. These Transformer-based models have two categories: Bi-encoders, which learn separate representations for the query and item; and Cross-encoders, which utilize attention to learn a joint representation of the query and item. By modeling full interaction between words of the query and item, Cross-encoders perform better than Bi-encoders. 
Cross-encoders have unbearable latency in practice since they cannot pre-compute the vector representations like Bi-encoders. 

 A common drawback of existing relevance models is that they can only provide predicted outputs and offer little transparency of their decisions. In contrast, a human can not only judge the relevance between the query and item but also illustrate the reason. For example, we annotate case (c) in Figure~\ref{fig:example} as relevance because the shopping intent of \emph{"gym weight"} matches the product type entity \emph{"dumbbell"}. It can be seen that human interpretation is often based on matching or mismatching of the query with entities in the item title. Hence, it is reasonable to formulate and interpret relevance based on entities. In the e-commerce scenario, all item titles consist of entities; it is reasonable to use entities as a form of the interpretable reason for prediction. Besides, when wrongly predicted results appear in the online search system, the bad cases need to be directly and efficiently intervened by human experts. However, in current search systems, the bad cases can only be fixed individually, which is very labor-intensive. 
 
 In this paper, we introduce Entity-based Relevance Model (EBRM), containing a query-entity (QE) relevance module and a prediction module using soft-logic. EBRM has comparable performance with Cross-encoders and faster inference speed by caching entity-based rules. Meanwhile, EBRM can provide entity-level justification of its query-item (QI) prediction. With interpretable entity-level justifications, the intervention of its prediction is efficient. The contributions of our work are: 
\begin{itemize}
    \item We propose a novel entity-based relevance model with desired properties for practical online search systems. (section \ref{sec:model})
    \item Training the QE relevance model is achieved through the indirect signal from the labeled QI relevance data. (section \ref{sec:inference_learning}) 
    \item We further propose a novel and effective method to pretrain EBRM with a massive amount of search logs, which significantly improves performance. (section \ref{sec:pretraining})
    \item Experimental results show that EBRM satisfies the desired properties of online serving: accurate, fast and memory-efficient, easy to interpret and intervene. (section \ref{sec:exp})
\end{itemize}

\section{Preliminary}
\paragraph{Problem Setup}
In this work, we refer to the user query as \textbf{query} and the product in the e-commerce system as \textbf{item}. For the product relevance prediction task, we first define the notations used in this paper. Let $ x = (Q, I)$ denote a pair of query text and item title text, $y$ be its relevance label. For this task, the relevance relation can be labeled into two classes: "relevant" and "not relevant", or more fine-grained classes to distinguish the subtle relevance difference \cite{yao2021learning, zou2021pre}, such as "strongly relevant", "relevant", "weakly relevant", "not relevant". In this work, we mainly focus on binary classes. 
We aim to build a binary classifier that takes $x = (Q, I)$ as input and predict whether the pair is relevant or not: $y = 1 $ if the pair $x$ is relevant or $y = 0$ otherwise.

\paragraph{Pretrained Language Model}
BERT \cite{devlin2019bert}, a pretrained language model with transformers \cite{NIPS2017_3f5ee243}, is trained on a huge amount of unlabelled text with Masked Language Model (MLM) and Next Sentence Prediction (NSP) losses. Recent works apply BERT to NER \cite{devlin2019bert, wang-etal-2020-pyramid, Fu_Tan_Chen_Huang_Huang_2021}, IR \cite{poly-encoder-2019, nogueira2019passage, cedr-2019, dai2019deeper}, syntactic parsing \cite{kulmizev-etal-2019-deep}, and achieve impressive performance. Several studies \cite{devlin2019bert, shi2019simple, peinelt-etal-2020-tbert} demonstrate that the representation of BERT could improve semantic tasks. 
\paragraph{Bi-encoder and Cross-encoder} \label{sec:bi-encoder}
Two common architectures for pairwise comparisons between texts are Bi-encoders and Cross-encoders. 
In a Bi-encoder, the query and item are encoded into two separate vectors:
\begin{align*}
\bar{h}_Q = \mathrm{pool}(\mathrm{T}(Q)) \qquad
\bar{h}_I = \mathrm{pool}(\mathrm{T}(I)) 
\end{align*}
where $\bar{h}_Q$ and $\bar{h}_I$ are contextualized representations of the query and item. $\mathrm{T}(\cdot)$ is a contextualized text encoder. A typical choice of the encoder is the pretrained BERT and the input text is in the form of "[CLS] <query text> [SEP]" and "[CLS] <item text> [SEP]". $\mathrm{T}(\cdot)=\{h_1, ..., h_n\}$ is the output of the transformer. $\mathrm{pool}(\cdot)$ is the pooling function. Following previous works, we choose the representation of the special token "[CLS]" as the representation of the whole input sequence.

To predict the relevance label of query-item pair $x$, a logistic classifier is used. In the Bi-encoder, the score is computed with a biaffine scoring function:
\begin{align*}
    P(y = 1 | Q, I) = \sigma(s(Q, I))  \\
    s(Q, I) = \bar{h}_Q^T W \bar{h}_I + b
\end{align*}
In the encoding procedure of the Bi-encoder, $\bar{h}_Q$ and $\bar{h}_I$ are encoded independently. As a result, one advantage of the Bi-encoder is that it allows us to precompute and cache the representations of queries and items, which accelerates the inference. This property enables the Bi-encoder to be deployed for online serving with low latency.

On the other hand, in a Cross-encoder, the query and title text are concatenated into a single text sequence and jointly encoded into a single vector. By using the transformer, the model obtains a representation with more interaction between the query and item than the ones from the Bi-encoder.
\[\bar{h}_{Q, I} = \mathrm{pool}(\mathrm{T}(\mathrm{concat}(Q, I)))\]
where $\mathrm{concat}(Q, I)$ is the function that concatenates the query text and the title text with the special token "[SEP]". Thus, the input text of the transformer is in the form of "[CLS] <query text> [SEP] <item text> [SEP]".

For the Cross-encoder, the score is computed with an MLP. 
\[s(Q, I) = \mathrm{MLP}(\bar{h}_{Q,I})\]
Since the representation can capture strong interaction between queries and items, the Cross-encoder could get better results than the Bi-encoder. However, precomputing and caching are impossible in the Cross-encoder. It has to re-encode every new (query, item) pair and hence cannot satisfy the need for low latency for online serving.

As we mentioned above, both Bi-encoder and Cross-encoder have their pros and cons and fail to satisfy all these desired properties for serving: accurate, rationalizable, intervenable and fast speed. As all these properties are essential for real-world application, we propose EBRM to reach a decent and well-balanced performance on all these pespectives, as shown in Table \ref{tab:model_property}, and empirically prove the comprehensiveness of EBRM in Section \ref{sec:exp}.

\begin{table}[t]
\setlength\tabcolsep{2pt}
    \small
    \centering
    \begin{tabular}{l c c c c}
         \toprule 
         & Accurate & Rationalizable & Intervenable & Speed \\
         \midrule
         \midrule
         QIRM \tiny{(Bi)} & \textcolor{red}{\Sadey[1.4]} & \textcolor{red}{\Sadey[1.4]} & \textcolor{red}{\Sadey[1.4]} & \textcolor{OliveGreen}{\Laughey[1.4]} \\
         QIRM \tiny{(Cross)} & \color{OliveGreen}{\Laughey[1.4]} & \textcolor{red}{\Sadey[1.4]} & \textcolor{red}{\Sadey[1.4]} & \textcolor{red}{\Sadey[1.4]} \\
         EBRM & \textcolor{OliveGreen}{\Laughey[1.4]}  & \textcolor{OliveGreen}{\Laughey[1.4]} & \textcolor{OliveGreen}{\Laughey[1.4]} & \textcolor{OliveGreen}{\Laughey[1.4]}  \\
         \midrule
        \bottomrule
    \end{tabular}
    \caption{Overall comparison of different models. }
    \label{tab:model_property}
\end{table}

\begin{figure}[t]
    \centering
     \includegraphics[width=0.45\textwidth]{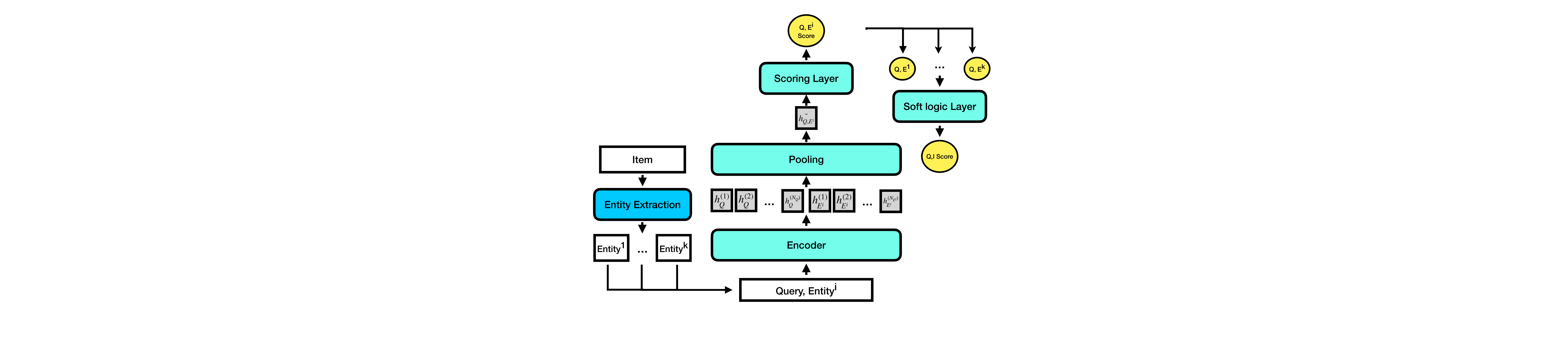}
     \caption{The diagram of entity-based relevance model. The query item relevance scoring is decomposed into several independent query entity scoring  and aggregated by a soft logic layer accordingly.  A Cross-encoder is utilized for fully exploring the interaction between the query and a specific entity. }
     \label{fig:ebrm}
\end{figure}

\section{Entity-based Search Relevance Model} 

The goal of general relevance prediction is to determine whether the query and the item are relevant or not. For e-commerce search, the product type is the most important entity type as all queries must have intended products to search for and all items must have intended products to sell. Thus, we focus on product type relevance prediction regardless of attribute mismatch in this work \footnote{Nonetheless, our method can be easily extended to general relevance prediction with some modifications, which we discuss in the Section \ref{sec:discussion}}. 

\subsection{Model} \label{sec:model}
Item titles are often verbose and contain more information than user queries. As a commonly used information extraction method, NER locates and classifies entity mentions in an unstructured text. We apply an off-the-shelf e-commerce NER model to extract entities from the item title. The extraction allows us to decompose the prediction procedure and offers us interpretable justifications for predictions, as will be explained below.\footnote{Entity extraction is seldom utilized in previous research on product search relevance models \cite{jiang2019unified, yao2021learning}. To make fair comparisons, in our experiments, we add comparisons with designed baselines that utilize NER prediction results.} The overall architecture is shown in Figure \ref{fig:ebrm}.

The e-commerce NER result of a title is a bag of entities $E_I = \{e_{1}, ... , e_{k}\}$. Focusing on product type relevance prediction, we only keep product type entity in the NER result and discard the others, such as brand entities. For example, the product type entity in the item "Intel Xeon Processor" is "Processor". "Intel Xeon" is the brand entity.

We follow the insight that a query and an item are relevant if and only if there exist some product type entities of the item that are relevant to the query. If a product type entity is a synonym or a hyponym of the query intent, it is relevant to the query. For example, a product type entity \emph{``phone cover''} is relevant to the query \emph{``phone case''}, because they are synonymous, while \emph{``case''} is irrelevant to the query because it is more general and is a hypernym of the query.

\[
\mathrm{Relevant}(Q, I) \iff \bigvee_{e \in E_I}{\mathrm{Relevant}(Q, e)}
\]

Our EBRM consists of a Cross-encoder QE relevance module and a soft logic aggregation layer for the final QI relevance prediction. The Cross-encoder QE relevance module uses one pretrained transformer to jointly encode the query and entity into a single vector.
\[\bar{h}_{Q, e} = \mathrm{pool}(\mathrm{T}(\mathrm{concat}(Q, e)))\]
For every QE (query, entity) pair in a QI (query, item) pair, the relevance probability is computed with an MLP scoring layer followed by a sigmoid function.
\begin{align*}
P(r = 1 | Q, e) &= \sigma(S(Q,e)) = \sigma(\mathrm{MLP}(\bar{h}_{Q, e}))
\end{align*}
where $r$ is the relevance label of the query and product type entity.

To predict the relevance label of the query and item, we regard the probabilities computed by the QE relevance module as soft matching of QE and aggregate them into soft matching of QI by applying a soft logic operator. Using the Zadeh soft logic \cite{ZADEH1965338}, we use max to replace disjunction and obtain the following formula.
\begin{align*}
     & P(y=1 | Q, I)  = \max_{e \in E_I}{P(r=1|Q, e)} \\
     &= \max_{e \in E_I}{\sigma(S(Q, e))} 
                     = \sigma(\max_{e \in E_I}{S(Q, e)})
\end{align*}
From the derivation, we define the score of QI as
\begin{align*}
    S(Q, I) = \max_{e \in E_I}{S(Q, e)}
\end{align*}

\subsection{Inference \& Learning} \label{sec:inference_learning}

The binary QI relevance prediction is obtained by
\begin{align*}
    y &= \arg\max_{y' \in \{0, 1\}}{P(y' | Q, I)} = [S(Q, I) \geq 0] \\
      &= [\max_{e \in E_I}{S(Q, e)} \geq 0] = \max_{e \in E_I}{[S(Q, e) \geq 0]}
\end{align*}
where $[\cdot]$ is Iverson bracket.
It can be seen that the QI relevance prediction is divided into several QE relevance predictions. The results of QE model act as direct explanations and signals for the QI predictions.

In practice, we can cache the QE relevance prediction results together with the entity recognition results of items and use them as rules for fast and accurate online inference, as we explained in Section~\ref{sec:serving} in Appendix.

Given a dataset $D=\{(x_1, y_1),  ...,  (x_n,y_n) \}$,  our training objective is to minimize the commonly-used negative log-likelihood to learn the parameters $\theta$. $x_i$ denotes a QI pair and $y_i$ denotes the relevance label.
\[
\theta^* = \arg\min_{\theta}{\frac{1}{n}\sum_{(x_i,y_i) \in D}{-\log P(y=y_i|x_i)}}
\]
From the perspective of QE relevance learning, the supervision signals come from the labeled QI data.

\subsection{Warm-up Pre-Training} \label{sec:pretraining}

\paragraph{QI Pretraining with Search Log}
The QE relevance module is the core component of our model. As mentioned above, we would train the QE relevance module with indirect supervision from QI data. However, such human-annotated QI data is laborious. At the same time, a e-commerce platform creates millions of user logs every day. The user-behavioral signals like click or purchase can be used for model learning.
Therefore, we propose to collect large-scale pseudo-labeled QI pairs from search logs according to user behavior. Given one query, we collect all its exposure items and count numbers of their clicks on the e-commerce platform in the past two months. We collect the QI pairs as positive examples whose numbers of clicks are the top $N$ among all QI pairs and take $M$ random QI pairs as negative examples if their exposure is more than $K$ and they are never clicked.

\paragraph{QE Pretraining with Search Log} Directly utilizing the click-through QI data in e-commerce might be misleading. Previous work \cite{DBLP:conf/sigir/ZhangWMH19,yao2021learning} find that the clicked results might be attributed to many factors including price, attractive titles or images\footnote{For example, during data collection, we find that some items are displayed many times but never or seldom clicked although the items are relevant to the query. This may be because these items do not have eye-catching photos.}. Therefore, we design a different approach that gathers QE pairs which can alleviate the mismatch problem in a single QI pair \footnote{An example of the procedure are shown in the Figure \ref{fig:qe_process} in the Appendix.}. Firstly, we collect all exposure items and count numbers of their clicks on the e-commerce platform given one query in the past two months. Then, we extract entities from the collected items and compute the number of each entity's click by summing the click numbers of all items which contain this entity. Lastly, we collect the QE pairs as positive examples whose click numbers are the top $N$ among all QE pairs and take the lowest $M$ pairs as negative examples. We construct a pseudo-labeled QE dataset in such a way and train our QE module on the pseudo-labeled dataset. Compared with QI relevance data, the QE training data is more useful to our QE module because the QE data can be seen as a distillation of the large-scale QI data that can be less noisy and more accurate. And the QE label is a direct training signal for our QE module. In our experiments, we set $N$ to 3, $M$ to 10 and $K$ to 10 by default.

\paragraph{QI Pretraining without Search Log}
For scenario where we cannot collect search logs from the same domain of the datasets, we propose to utilize the parameters of a well-trained Cross-Encoder model, which is named as "QIRM (Cross)" in Section \ref{sec:approach}, to initialize the parameters of the EBRM model.

\subsection{Properties of EBRM} 

Our proposed entity-based relevance model utilizes entities to bridge the gap between queries and items. Here we discuss why our entity-based relevance model satisfies some of the properties list in Table \ref{tab:model_property}. Empirical results verifying these properties are shown in Section \ref{sec:exp}.

\paragraph{Rationalizable} Since our model uses a soft logic aggregation layer for final prediction, the logic layer allows us to interpret every prediction result of a QI pair at the entity level. If the predicted result of a query and an item is relevant, the cached entities for the query and item should have at least one overlapping entity, which can be regarded as the explanation of the relevance prediction.

\paragraph{Intervenable} Since the prediction procedure is rationalizable, if we find a critical error in the QE relevance rule, we can hotfix the error to improve the performance of the relevance module. Specifically, we can add or delete a specific entity for each query to change the search results. By intervening one specific QE relevance rule, we could influence the predictions of the query and all items having this entity, which is efficiency.

\paragraph{Speed} With precomputation and caching the QE relevance rules, the inference procedure only needs simple and fast entity matching, that is, the model only needs to check whether entities in QE rules appear in the item. The speed of our model could be comparable with Bi-encoders.

\section{Experiments} \label{sec:exp}
To verify the advantages of our model, we conduct a variety of experiments with different datasets.

\subsection{Datasets}
The experiments are conducted on a private dataset and an open-source dataset. For the private dataset, our training data is collected from the search logs of an e-commerce website. For the QI dataset, the relevance labels are annotated by humans. The train-split is used for model training and the dev-split is used for model selection. As mentioned above, we automatically generate a large amount of pseudo-labeled QI and QE data for pretraining. 
Meanwhile, to enable reproducible research, we also utilize a recently released open-source e-commerce dataset WANDS \cite{wands} \footnote{According to its annotation guidelines, we treat the labels "exact match" and "partial match" as "product type relevant" and "irrelevant" as "product type irrelevant". We split the whole WANDS dataset into train, dev, test sets in 3:1:1 ratio. The split dataset will be released in our git project.}.  
The statistics of datasets is shown in Table~\ref{tab:dataset_statistic} in Appendix.

\begin{table}[]
    \scriptsize
    \centering
    \begin{tabular}{@{\hspace*{2em}} l c c  c c}
    \toprule 
    & \multicolumn{2}{c}{\textbf{E-commerce}} & \multicolumn{2}{c}{\textbf{WANDS}} \\
     & \textbf{Acc}  & \textbf{F1}  & \textbf{Acc} & \textbf{F1}  \\
    \midrule
    \midrule
    
    \multicolumn{5}{l}{\textbf{With Cross-Encoder: for reference}} \\ 
    QIRM (Cross) &	86.58 	&		72.73 	&	89.37 	 &	86.03  \\
    QEsRM (Cross) &	85.51 	&	    68.43 	&	87.40 	 &	84.34  \\

    \midrule
    \multicolumn{5}{l}{\textbf{Existing Text matching models}} \\
    ARC-II \cite{hu2014convolutional}  & 82.58  & 57.64 & 54.41  &	49.45   \\
    CDSSM  \cite{shen2014learning} & 83.04    & 50.27  & 34.10 & 28.57  \\
    MatchPyramid \cite{pang2016text} & 83.49   & 57.81  & 62.11  & 52.77  \\
    KNRM \cite{KNRM-2017} & 82.35    & 57.12  & 60.89  & 53.27  \\
    ConvKNRM \cite{KNRM-2018} & 83.64  & 61.37  & 70.33  & 57.01  \\
    
    \midrule
    \multicolumn{5}{l}{\textbf{Entity Recognition (Aug by KBs)}} \\
    NER & 76.68  & 64.21 & 54.96 & 52.95 \\
    NER Aug w/ Syn & 77.72  & 64.85 & 52.92 & 51.66 \\
    NER Aug w/ Syn \& Related Es & 80.60  & 65.20 &  53.01& 52.88\\
    \midrule 
    \multicolumn{5}{l}{\textbf{With Bi-Encoder}} \\ 
    QIRM (Bi) &	84.13 	&	61.29 	&	61.88 		&	59.26   \\
    QEsRM (Bi) &	83.69 		&	59.52 	&	65.17 	&	61.40  \\
    QIRM (Bi) + QI pretrain &	84.98  		&	63.82 	&	-	&	- \\
    \midrule \midrule
    \multicolumn{5}{l}{\textbf{Entity-based relevance model: this work}} \\
    EBRM & 84.84 	&	66.37 	&	 87.45 	 	&	 83.59  \\
    EBRM + QE pretrain  &	\bf 87.25 	 	&	\bf 71.98 	&	-		&	- \\
    EBRM + Cross Init & 84.72 	 & 68.53 	& \bf 87.62 & \bf 84.16 \\

    \bottomrule
    \end{tabular}
    \caption{Performances on QI relevance datasets. }
    \label{tab:main_performance}
\end{table}

\subsection{Compared Approaches}
\label{sec:approach}
We organize the approaches into five groups:

\subsubsection{Existing Text Matching Models}
We conduct experiments with several existing neural text matching models: Arc-II \cite{hu2014convolutional}, CDSSM \cite{shen2014learning}, MatchPyramid \cite{pang2016text}, KNRM \cite{KNRM-2017} and ConvKNRM \cite{KNRM-2018}.

\subsubsection{Entity Recognition (Aug by KBs)}

\paragraph{Relevance Prediction with Pure NER}
We extract entities from the query and item in each QI pair by an off-the-shelf NER tool respectively, and then we match these extracted entity texts. If there exists one or more entities appearing in both the query and the item, we predict the QI as relevant; otherwise, we predict it as irrelevant. If a query does not have a product type entity, we regard all the exposed items to the query as relevant.

\paragraph{Relevance Prediction with Pure NER \& Existing Knowledge Base} To alleviate the surface form gap between the query and item, we try to enrich the information of the query based on the publicly released knowledge base ConceptNet \cite{conceptnet}. We can expand the entities with relation "Synonym" (referred as \texttt{NER Aug w/ Syn}) or the "Synonym", "SimilarTo" and "RelatedTo" relations simultaneously (referred as \texttt{NER Aug w/ Syn \& Related Es}) 
ConceptNet has triplets $\langle\textsc{ent}_1, \textsc{relation}, \textsc{ent}_2 \rangle$.
Following \cite{majilogic}, we take the entity extracted from the query as $\textsc{ent}_1$ and obtain all $\textsc{ent}_2$ as the expanded entity while the relation between these two entities is one of "Synonym", "SimilarTo" and "RelatedTo". In such way, we can extend structured information of the query representation.
We compare two settings while their difference is which kinds of relations we take into consideration. 
After obtaining the synonym or related entities of the extracted entities from the query
, we match the original and expanded entities from the query and the entities from the title in the same way as in the "Relevance Prediction with Pure NER" approach.

\subsubsection{Bi-encoder Models}

\paragraph{QIRM (Bi)} One important baseline model is the Bi-encoder model, which is described in Section \ref{sec:bi-encoder}. The Bi-encoder models have been widely used in previous work \cite{DBLP:conf/kdd/0006RHMGLB21,DBLP:conf/kdd/HuangSSXZPPOY20}, which have been shown to be a strong baseline.  For QIRM (Bi), the inputs are the query and item title.  We also enhance it by QI pretraining. 

\paragraph{QEsRM (Bi)} To analyze the influence of entity extraction, we modify the inputs of the Bi-encoder. Specifically, the item title is replaced with a concatenated product type entities text.

\subsubsection{Cross-encoder Models for reference}

\paragraph{QIRM (Cross)} The input of the Cross-encoder is the concatenation of the query and item title. 

\paragraph{QEsRM (Cross)} Similar to the QEsRM (Bi) baseline, but with concatenated product type entity texts as the input of the item.

\subsubsection{Our proposed Entity-based Models}

\paragraph{EBRM}
Our proposed entity-based relevant model (EBRM) is trained with the labeled QI data. We enhance it with QE pretraining and initialization with Cross-encoder.

For EBRMs, QIRMs and QEsRMs, we use BERT-base-uncased as the encoder\footnote{Since BERT uses the NSP objective in pretraining and uses different embeddings for different segments, we set the segment of the query to 0 and the other to 1. When using the Bi-encoder, the input of the transformer is single text. We set the segment of input to 0.}.

\subsection{Results on QI Relevance}
For evaluation, we use the following metrics: Accuracy, macro F1 score. The experimental results are shown in Table \ref{tab:main_performance}.\footnote{Please refer to the Table \ref{tab:tab:performance_detailed} in Appendix for detailed performance.}

\emph{Observation \#1}: By capturing deep interaction between queries and items, the Cross-encoders have better performance than the Bi-encoders as expected. Compared with the Bi-encoder baselines QIRM (Bi) and QEsRM (Bi), our EBRM outperforms them on accuracy and F1 score, which demonstrates the effectiveness of our model.  The advantage over QEsRM (Bi) also verifies that the performance gain does not completely come from the entity extraction model.

\emph{Observation \#2}: On E-commerce dataset, with the help of QE module, EBRM has substantial improvement in performance and is comparable with the QIRM (Cross). We think the reason is that the QE pseudo-labeled data provides clean negative samples for the QE module, which improves its discriminative ability. On both datasets, initialization with Cross-encoder parameters can slightly improve the performance of EBRM.

\emph{Observation \#3}: 
For both datasets, We can find that the entity recognition methods achieve much poorer overall accuracy than our proposed EBRM. For E-commerce dataset, we find the entity recognition methods have an extremely poor balance between accuracy of positive and negative samples.
This is due to that the gap between the surface forms of the query and item is huge, so that entities in relevant query and title pairs are often different (although their semantic meanings are similar).
With the help of the public knowledge base ConceptNet, we can reduce the gap. 
However, such entity recognition methods have an obvious disadvantage that the coverage of entity augmentation for queries is low, because lots of queries may not have product type entities and hence cannot be easily expanded. Besides, the publicly released knowledge base may not match the e-commerce scenario well while building an accurate and high-coverage e-commerce knowledge base is difficult.

\emph{Observation \#4}: 
Compared with the classic neural text matching models, all the models (including QIRMs, QEsRMs, EBRMs) with pretrained language models perform better on all Acc and F1 metrics, which shows the effectiveness of pretrained language models.

\subsection{Intervention Ability}
We simulate the intervention process of EBRM on our collected pseudo-labeled QI data from user logs to investigate the intervention efficiency.
We randomly select 120 queries, and for each query we collect about 500 purchased items as positive examples and 100 items never been clicked as negative examples. 
We intervene our QI prediction by supplying/deleting entities to each query from its false negative/positive examples respectively. 
From Table \ref{tab:intervention}, we can find that our intervention process can improve the performance of QI prediction. Furthermore, for each QI intervention operation, our method only takes 0.26/1 $\approx$ 1/4 actions, which is much faster than the naive QI intervention method.

\subsection{Inference Speed}

\begin{table}
    \tiny
    \setlength\tabcolsep{4pt}
    \centering
    \scalebox{1.2}{
    \begin{tabular}{l|c c c}
         \toprule 
         & \tiny{Accuracy} & \tiny{\makecell{Speed  \\ (\#Instance/s)}} & \tiny{\makecell{Extra Cache \\ Space (MB)}}  \\
         \midrule
         EBRM \tiny{+ QE pretrain}  &  \bf 87.25 & \bf 4898.50   & \bf 0.5   \\
         QIRM \tiny{(Bi)}           &  84.13      & 2211.07       & 500     \\
         QIRM \tiny{(Cross) [BERT-base]}  &  86.58   & 4.38       & -       \\
         QIRM \tiny{(Cross) [BERT-small]} &  85.71   & 5.42       & -       \\
         QIRM \tiny{(Cross) [BERT-tiny]}  &  85.23   & 9.37       & -       \\
         \bottomrule
    \end{tabular}}
    \caption{Results on speed comparison experiments.}
    \label{tab:speed}
\end{table}

As we mentioned before, one advantage of EBRM is its reasonable inference speed. We compare with the Cross-encoder model and the Bi-encoder model. For the Cross-encoder, we further test the speed of two small variants with BERT-small-uncased and BERT-tiny-uncased.
For the Bi-encoder, we obtain the vectors of queries and items in the experimental data from a trained Bi-encoder. For EBRM, we precompute with the QE relevance module and cache the QE relevance rules.\footnote{In our statistics of a real anonymous ecommerce website, we can cache over 80 percent of user queries that previously searched.} All experiments are conducted with the test set of the E-commerce dataset.

Experiment results are presented in Table \ref{tab:speed}. With smaller BERT variants, the Cross-encoders become faster but less accurate. With the help of precomputation, EBRM is much faster than QIRM (Cross) with different BERT variants. EBRM has a slight advantage in speed over QIRM (Bi) since its simpler computation. Besides, EBRM only needs to cache the sparse QE relevance rules, which is more memory-efficient than caching vectors from QIRM (Bi). These results show that our model is suitable for practical applications from the perspective of efficiency.   

\begin{table}
    \small
    \centering
    \begin{tabular}{l|c c}
         \toprule 
         & Accurate & \#Action/QI  \\
         \midrule
         \midrule
         prev     &  83.88        & -   \\
         post     &  89.36        & 0.26   \\
         \bottomrule
    \end{tabular}
    \caption{Results on intervention experiments. Actions means the modifications of QE/QI predictions.}
    \label{tab:intervention}
\end{table}

\section{Discussion: Extending to General Relevance Prediction}
\label{sec:discussion}
In this paper, we focus on product type relevance prediction. For the general QI relevance prediction problem, we make modifications to the input processing and the soft logic layer of the model. In general, there are many other entity types except product, such as Brand and Color. We apply a NER model to extract the entities of the query. Our belief of general relevance is: for each entity type $c$ that occurs in the query, a matched item needs to have a relevant entity of this type.
\begin{displaymath}
\mathrm{Relevant}(Q, I) \iff \bigwedge_{c \in C_Q}{\bigvee_{e \in E_I^c}{\mathrm{Relevant}(Q, e)}}
\end{displaymath}
where $C_Q$ is the set of entity types appearing in the query and ${E_I^c}$ is the entity set of type $c$ in the item.

\section{Related Work}

\subsection{Search Relevance Models}
The relevance model is an important part of a search system and is used in retrieval and ranking \cite{dssm-2013, DRMM-16, KNRM-2018}. Recent studies \cite{nogueira2019passage, cedr-2019, dai2019deeper} utilize pretrained language models, notably BERT \cite{devlin2019bert}. However, PLM-based relevance models are time-consuming for online serving in practice. In order to narrow the gap, some works redesign the architecture for speedup \cite{colbert-2020, poly-encoder-2019}. Compared with these approaches, we focus on the relevance model for e-commerce product search and propose a novel entity-based model architecture. It can employ the expensive PLM-based relevance computation offline and provide a speedup for online prediction.

\subsection{Entity-Oriented Search}
Entities are semantic units for organizing information, which can be utilized as strong relevance ranking features \cite{balog2018entity}. For example, entity-based information can be utilized in improving language models \cite{DBLP:conf/sigir/RavivKC16}. 
Another application is entity search \cite{balog2010entity}, which aims to retrieve relevant items from a semantic data set about entities. Compared with these studies, we propose to bridge the gap between queries and items with product type entities. Furthermore, our proposed approach explicitly utilizes entities to directly compute the relevance results.

\subsection{Marrying Neural \& Symbolic Methods}
Several recent research works seek to integrate symbolic knowledge (such as logic rules) into neural networks. One direction is integrating logic rules into training through different methods, such as posterior regularization \cite{hu-etal-2016-harnessing}, consistency loss \cite{li-etal-2019-logic} and discrepancy loss \cite{wang2020integrating} between neural networks and logic, construction of adversarial sets \cite{minervini-riedel-2018-adversarially}. 
Another direction directly augments neural networks with logic neurons \cite{li-srikumar-2019-augmenting}.  Motivated by these studies, our method integrates the logic rules of product relevance matching as a soft logic layer in the neural network and retains strong interpretability of prediction from the logic rules. 

\section{Conclusion}
For solving the product type relevance prediction problem in e-commerce search, we propose a novel relevance model called EBRM. EBRM decomposes the QI prediction problem into several QE prediction problems and aggregates the results with a soft logic module. From the perspective of online serving, EBRM has advantages over the commonly-used Bi-encoders and Cross-encoders and is rationalizable, intervenable, fast and accurate.  Experimental results verify that our model is effective and beneficial for practical applications. 

\bibliography{anthology,custom}
\bibliographystyle{acl_natbib}

\newpage 
\mbox{}
\newpage

\appendix

\section{Serving EBRM Online} \label{sec:serving}

Similar to previous work on caching vectors for both queries and items \cite{DBLP:conf/kdd/0006RHMGLB21,DBLP:conf/kdd/HuangSSXZPPOY20}, we propose to cache the tagged entities for both queries and items. During the online serving period, we check whether a specific query and item pair has at least one overlapping entity to produce the relevance result.

\paragraph{Caching Entities for Query} 
To obtain the candidate relevant entities of each query and reduce the computation of QE prediction, we design a data collection method containing five steps. The previous three steps are used to collect candidate QE pairs and the procedure is same as above mentioned in Section 3.3. Based on the collected QE pair and their numbers of clicks, we take two more steps. 1). We choose the entities as candidate entities whose numbers of clicks are the top 100 for each query. 2). We use our proposed model to predict whether each candidate QE pair is relevant.  Such procedures are performed in an offline manner. Finally, we store the entities relevant to the query for online prediction \footnote{In our statistics of a real anonymous ecommerce website, we can cache over 80 percent of user queries that previously searched.}.
\paragraph{Caching Entities for Item} We utilize an off-the-shelf entity extraction tool to directly recognize the product type entities in the item title and store them respectively.

\section{Case study}
\begin{figure}[t]
    \centering
    \scalebox{0.23}{
     \includegraphics{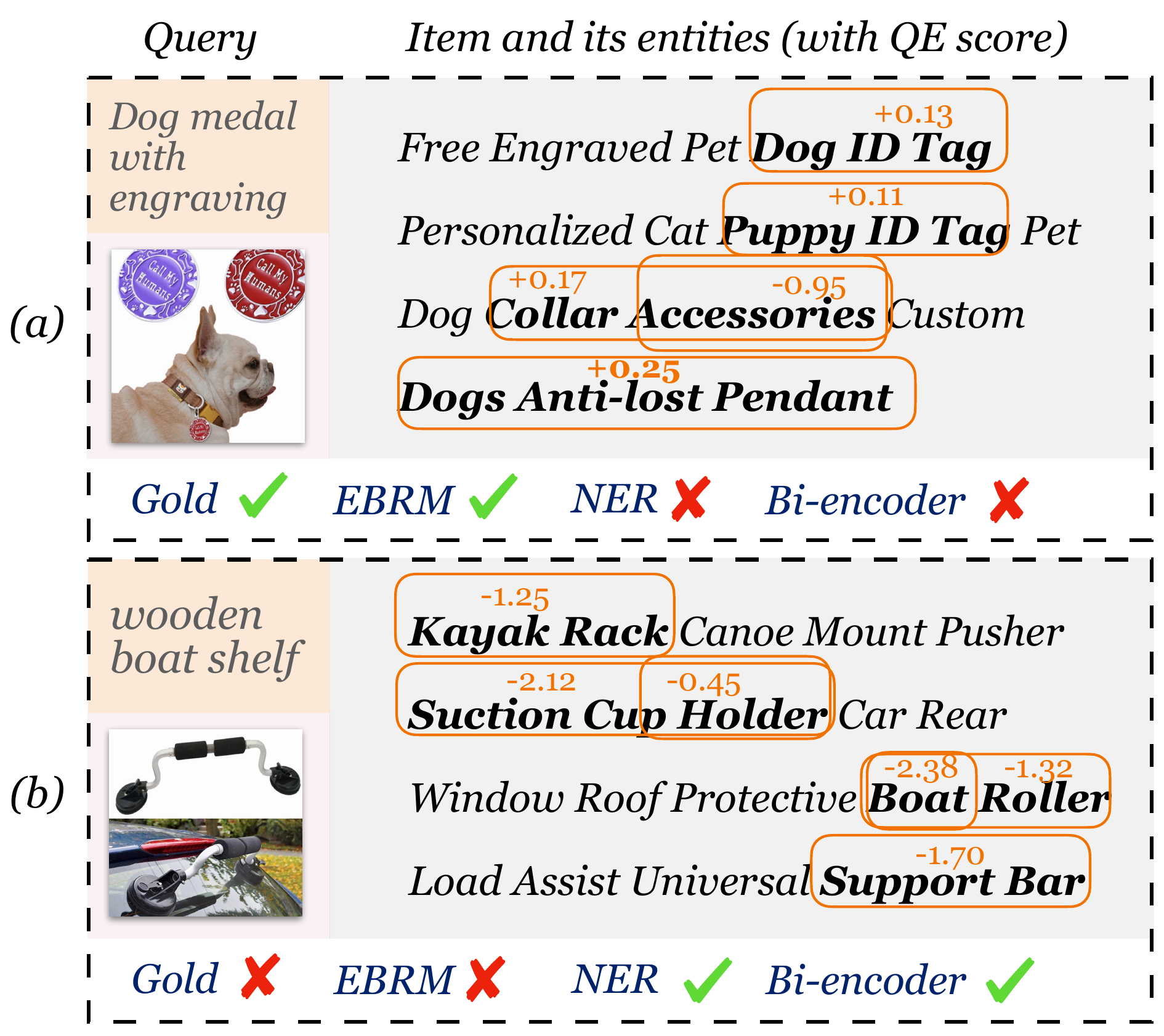}
    }
    \caption{Case study. \cmark\  means the gold or predicted label of QI pair is relevant and \xmark\  means they are irrelevant.}
    \label{fig:casestudy}
\end{figure}

In this section, we analyze the difference between models as shown in Figure \ref{fig:casestudy}. From these two examples, we find that our model EBRM correctly tags the query with accurate entities compared with the NER methods. Compared with the Bi-encoder, EBRM not only precisely predicts relevance output, but also offers entity-level explanations. 
In example (a), EBRM finds that entities from items, including ``\emph{Dog ID Tag}'', ``\emph{Collar Accessories}'', ``\emph{Dog Anti-lost Pendant}', are relevant to the query, and hence predict the QI pair to be relevant. However, due to different surface forms of dog medals between the query and item, the NER methods failed to make the right prediction. Besides, we cannot easily analyze why Bi-encoder makes erroneous predictions for a specific case.
In example (b), EBRM regards all entities in the title as irrelevant to the query and therefore predict the label as irrelevant,  but Bi-encoder makes a mistake again. 


\section{Experiments: Online Evaluation} 
\label{appendix: online_eval}
We deployed the proposed model online in the search system of a real anonymous e-commerce website. Compared with the online search relevance model as the baseline, we perform online A/B testing to investigate the effectiveness of EBRM. Both experiments take about 10\% proportion of search traffic. To evaluate the empirical results on search relevance, we ask two human annotators to label the search results of randomly sampled queries over the website. If these two annotators produce different annotations for a QI pair, we will ask a third annotator to label the QI. We find that the annotation consistency is 95\%. Motivated by previous work on online search relevance evaluation \cite{yao2021learning,zou2021pre}, we evaluate the ratio of relevant QI for both the base and test buckets. Results show that our proposed method outperforms the online search system and improves the search relevance of the website by 1.29\%. 
Besides, we find that the scale of queries or items is much bigger than that of entities as shown in Table \ref{tab:dataset_statistic_online}.  Entities per query or item is not more than 10, which means that our proposed EBRM has a small memory storage cost, which is important for online e-commerce systems.

\begin{table*}[t]
    \small
    \centering
    \begin{tabular}{c|c|c|c|c|c|c}
    \toprule
        Split & \#Sample & \#Query / Avg Len & \#Item / Avg Len & \#Entity/ Avg Len & \#Relevant & \#Irrelevant \\
    \midrule
    \midrule
        \multicolumn{7}{c}{E-commerce: QI} \\
    \midrule
    Pre-Train$^\dagger$ & 10,000,000 & 841,058 / 2.92 & 4,774,587 / 18.16 &	N/A	 & 1,792,077 & 8,207,923 \\
    Train & 248,156 & 31,564 / 2.71 & 128,510 / 18.19 & 95,700 / 2.33 & 204,763 & 43,393 \\
    Dev & 36,508 & 4,626 / 2.29 & 29,827 / 18.03 & 29,683 / 2.2 & 29,138 & 7,370 \\
    Test & 15,497 & 1,614 / 2.92 & 15,163 / 18.46 & 18,480 / 2.15 & 12,964 & 2,533 \\
    \midrule
        \multicolumn{7}{c}{E-commerce: QE} \\
    \midrule
    Pre-Train$^\dagger$ & 10,000,000 & 769,270 / 3.21 & N/A  & 2,055,948 / 2.53 & 2,307,804 & 7,692,196 \\
    \midrule
        \multicolumn{7}{c}{WANDS} \\
    \midrule
    Train & 53,375 & 273 / 3.34 & 28,994 / 6.95 & 9,238 / 2.2 & 14,772 & 38,603 \\
    Dev & 18,261 & 91 / 3.3 & 14,416 / 7.55 & 4,760 / 2.1 & 6,709 & 11,552 \\ 
    Test & 14,765 & 91 / 3.57 & 11,957 / 7.72 & 4,632 / 2.15 & 4,123 & 10,642 \\

    \bottomrule
    
    \end{tabular}
    \caption{Statistics of relevance datasets. $\dagger$ marks the generated pseudo-labeled dataset.}
    
    \label{tab:dataset_statistic}
    \vspace{-2mm}
\end{table*}

\begin{table}[t]
    \small
    \centering
    \begin{tabular}{c|c|c|c|c}
    \toprule
         & \#Sample & \#Entity & \#Es/Sample & \#Unique Es  \\
    \midrule
    Query & 9,630,854 & 92,653,475 & 9.62 & 402,046  \\
    Item & 165,136,387 & 641,444,980 & 3.88 & 402,046 \\
    \bottomrule
    
    \end{tabular}
    \caption{Statistics of items and queries from the online serving system. Es is regarded as entities for short. }
    \label{tab:dataset_statistic_online}
    \vspace{-4mm}
\end{table}

\begin{table*}[t]
    \tiny
    \centering
    \begin{tabular}{@{\hspace*{2em}} l c c c c  c c c c}
    \toprule 
    & \multicolumn{4}{c}{\textbf{E-commerce}} & \multicolumn{4}{c}{\textbf{WANDS}} \\
     & \textbf{Acc} & \textbf{PosAcc} & \textbf{NegAcc} & \textbf{F1}  & \textbf{Acc} & \textbf{PosAcc} & \textbf{NegAcc} & \textbf{F1}  \\
    \midrule
    \midrule
    \multicolumn{1}{l}{\textbf{With Cross-Encoder: for reference}} \\ 
    QIRM (Cross) &	86.58 $\pm$ 0.16	&	94.34 $\pm$ 0.38	&	46.88 $\pm$ 1.40	&	72.73 $\pm$ 0.35	&	89.37 $\pm$ 1.57	&	72.73 $\pm$ 5.33	&	95.81 $\pm$ 2.30	&	86.03 $\pm$ 2.05 \\
    QEsRM (Cross) &	85.51 $\pm$ 0.60	&	95.02 $\pm$ 1.69	&	36.84 $\pm$ 5.13	&	68.43 $\pm$ 1.16	&	87.40 $\pm$ 0.38	&	77.46 $\pm$ 2.59	&	91.25 $\pm$ 1.24	&	84.34 $\pm$ 0.43 \\

    \midrule
    \multicolumn{1}{l}{\textbf{Existing Text matching models}} \\
    ARC-II \cite{hu2014convolutional} &	82.58 $\pm$ 1.79	&	95.12 $\pm$ 3.45	&	18.41 $\pm$ 6.67	&	57.64 $\pm$ 1.67	&	54.41 $\pm$ 2.59	&	41.50 $\pm$ 2.46	&	59.42 $\pm$ 4.50	&	49.45 $\pm$ 1.56  \\
    CDSSM \cite{shen2014learning} & 83.04 $\pm$ 0.87 & 97.87 $\pm$ 3.12 & 7.11 $\pm$ 10.65 & 50.27 $\pm$ 6.69 & 34.10 $\pm$ 12.34& 86.33 $\pm$ 27.35& 13.86 $\pm$ 27.72& 28.57 $\pm$ 13.48 \\
    MatchPyramid \cite{pang2016text} & 83.49 $\pm$ 1.15 & 96.48 $\pm$ 2.23 & 17.00 $\pm$ 4.58  & 57.81 $\pm$ 1.54 & 62.11 $\pm$ 3.42& 31.79 $\pm$ 3.95& 73.86 $\pm$ 6.03& 52.77 $\pm$ 1.70 \\
    KNRM \cite{KNRM-2017} & 82.35 $\pm$ 0.79 & 95.01 $\pm$ 1.81 & 17.58 $\pm$ 5.46 & 57.12 $\pm$ 2.39 & 60.89 $\pm$ 3.44& 36.74 $\pm$ 1.20& 70.24 $\pm$ 5.22& 53.27 $\pm$ 2.27 \\
    ConvKNRM \cite{KNRM-2018} & 83.64 $\pm$ 0.76 & 95.37 $\pm$ 0.53 & 23.59 $\pm$ 2.09 & 61.37 $\pm$ 1.68 & 70.33 $\pm$ 1.95& 26.45 $\pm$ 3.60& 87.33 $\pm$ 4.10& 57.01 $\pm$ 0.14 \\
    
    \midrule
    \multicolumn{1}{l}{\textbf{Entity Recognition (Aug by KBs)}} \\
    NER & 76.68 & 81.11 & 54.00 & 64.21 & 54.96& 61.36& 52.48& 52.95 \\
    NER Aug w/ Syn & 77.72 & 82.62 & 52.66 & 64.85 & 52.92& 65.90& 47.89& 51.66 \\
    NER Aug w/ Syn \& Related Es & 80.60 & 87.93 & 43.07 & 65.20 &  53.01& 85.59& 40.39& 52.88\\
    \midrule 
    \multicolumn{1}{l}{\textbf{With Bi-Encoder}} \\ 
    QIRM (Bi) &	84.13 $\pm$ 0.29	&	96.19 $\pm$ 0.25	&	22.40 $\pm$ 1.37	&	61.29 $\pm$ 0.88	&	61.88 $\pm$ 2.32	&	65.48 $\pm$ 6.04	&	60.49 $\pm$ 1.60	&	59.26 $\pm$ 2.65  \\
    
    QEsRM (Bi) &	83.69 $\pm$ 0.42	&	96.19 $\pm$ 0.96	&	19.72 $\pm$ 2.48	&	59.52 $\pm$ 0.94	&	65.17 $\pm$ 5.09	&	61.53 $\pm$ 7.65	&	66.58 $\pm$ 9.67	&	61.40 $\pm$ 3.45 \\
    
    QIRM (Bi) + QI pretrain &	84.98  $\pm$  0.33	&	96.49 $\pm$ 0.55	&	26.07 $\pm$ 2.24	&	63.82 $\pm$ 1.07	&	-	&	-	&	-	&	- \\
    \midrule \midrule
    \multicolumn{1}{l}{\textbf{Entity-based relevance model: this work}} \\
    EBRM & 84.84 $\pm$ 0.27	&	95.01 $\pm$ 0.50	&	32.84 $\pm$ 1.94	&	66.37 $\pm$ 0.75	&	87.45 $\pm$ 0.68	&	69.86 $\pm$ 4.31	&	94.27 $\pm$ 1.53	&	83.59 $\pm$ 1.06 \\
    EBRM + QE pretrain  &	87.25 $\pm$ 0.42	&	96.23 $\pm$ 0.38	&	41.23 $\pm$ 3.84	&	71.98 $\pm$ 1.64	&	-	&	-	&	-	&	- \\
    EBRM + Cross Init & 84.72 $\pm$ 0.24	& 93.49 $\pm$ 0.63	& 39.80 $\pm$ 2.25	& 68.53 $\pm$ 0.59	& 87.62 $\pm$ 1.34	& 73.15 $\pm$ 4.09	& 93.23 $\pm$ 2.47	& 84.16 $\pm$ 1.52 \\

    \bottomrule
    \end{tabular}
    \caption{Performances on QI relevance datasets. }
    \label{tab:performance_detailed}
\end{table*}

\begin{figure*}[t]
    \centering
    \scalebox{0.4}{
     \includegraphics{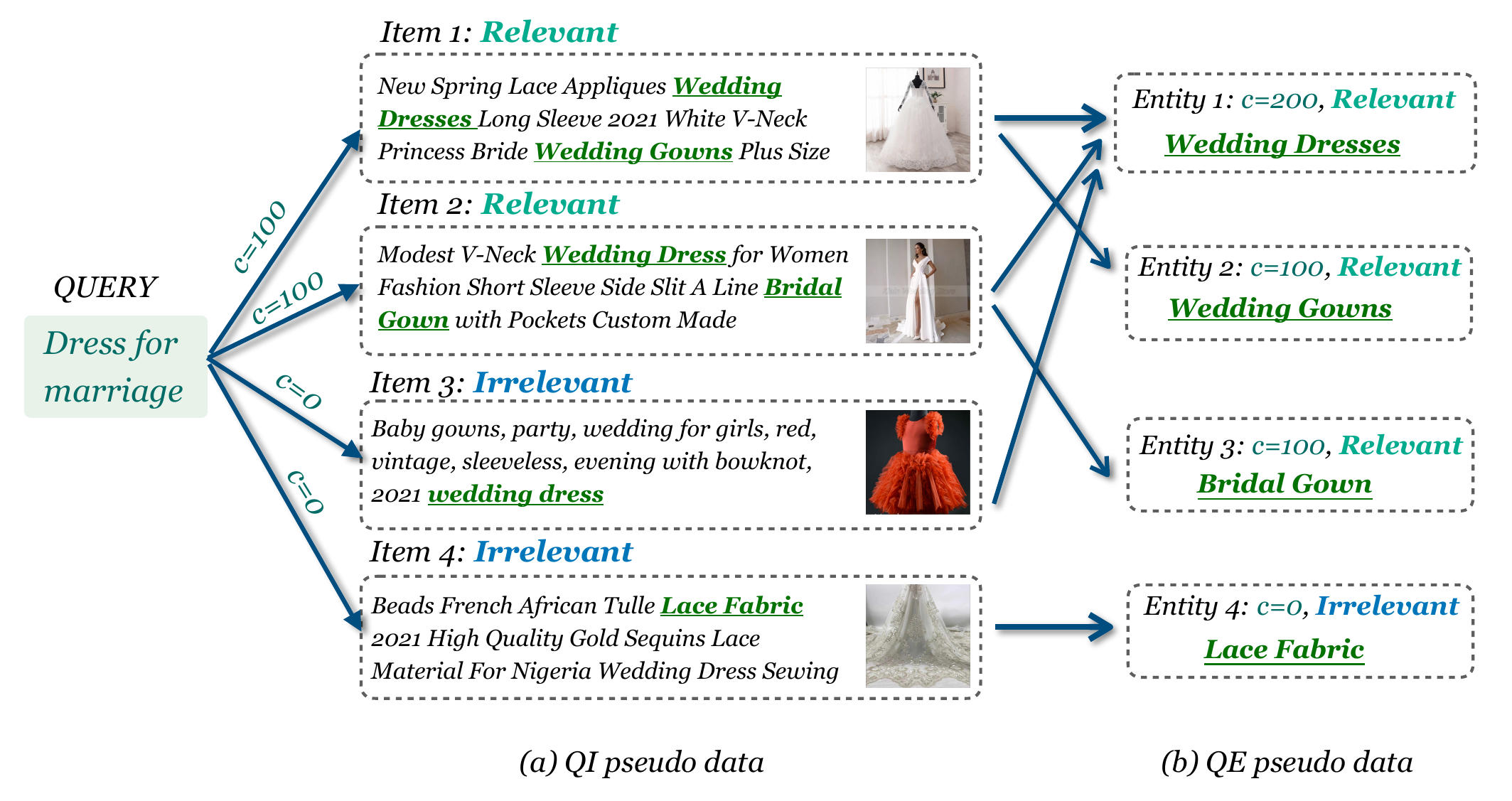}}
     \caption{The procedure of QI/QE pseudo-labeled data generation from a large-scale search log. To generate QI data, we collect all items which are listed by the search engine given a query in the past two months and count their numbers of clicks. To produce QE data, we take two more steps: 1). We extract entities from these exposed items and aggregate these entities. 2). We compute the click times of each entity by summing the click numbers of all items which contain the entity. We decide whether the QI/QE is relevant according to the click numbers. (We use $c$ to represent the  number of clicks between query and items/entities.) }
     \label{fig:qe_process}
\end{figure*}

\end{document}